# Resonance scattering on the lattice with non-zero total momentum

Steven Gottlieb and K. Rummukainen

Indiana University, Department of Physics, Swain Hall-West 117, Bloomington, IN 47405, USA

Most hadronic particles are resonances: for example, the $\rho$ meson appears as a resonance in the elastic scattering of two pions. A method by Lüscher enables one to measure the properties of the resonance particles from finite lattices. We present a more general method which includes scattering processes where the total momentum of the particles is non-zero. The main advantage is that the resonance scattering can be observed in a considerably smaller spatial volume. We test the method with a simple 3+1 dimensional spin model, and find excellent agreement between the zero momentum and the non-zero momentum scattering sectors.

## 1. INTRODUCTION

Lattice simulations have been very succesful in resolving many low energy properties of hadronic particles. However, many low energy hadrons are resonances, which cannot be fully described without scattering into asymptotic states. The best known example is the $\rho$ meson which appears in the elastic $\pi\pi$ $L=1$, $I=1$ scattering channel.

Elastic scattering involves at least four asymptotic states, which, in the case of QCD, are themselves bound states of fundamental quark and gluon fields. Because the lattice simulations are restricted to small (periodic) physical volumes, the direct observation of free asymptotic states is not possible. At best, the lattice size is a few times the interaction length.

This problem has been addressed in a series of papers by Lüscher [1], giving a relation between the infinite volume elastic scattering phase shift and the 2-particle energy spectrum measured on finite periodic lattices. The finite volume is used to probe the system, and simulations with several volumes are needed to obtain full knowledge of the phase shift function. This method has been applied to various test models in 2–4 dimensions [2–5].

Lüscher's method requires that the scattering particles have zero total momentum $\vec{P}=0$. We have generalized the method to *arbitary total momentum* $\vec{P}$ [6]. This method has several desirable features:

- When $\vec{P}\neq 0$ the resonance can be observed in smaller volumes than with $\vec{P}=0$.

- $\vec{P}=0$ and $\neq 0$ sectors can be measured simultaneously; depending on the model, often with negligible cost in computer time.

- For a given lattice size, the $\vec{P}=0$ and $\neq 0$ methods yield the phase shift $\delta_l(p)$ with a different argument $p$ – two data points from a single run.

To illustrate the difference in the required lattice size, consider a 1-dimensional system containing a light particle $\phi$, and a heavy resonance $\rho$. Let us first consider the energy levels in the *non-interacting* case. With lattice size $L$, the lattice momenta are quantized to $p = n\,2\pi/L$, $n\in\mathbf{Z}$. When we later turn on a small $\rho\phi\phi$ interaction, the resonance appears when the energies of the $\rho$ and $2\phi$-states are equal: $W_\rho = W_{2\phi}$. For concreteness, let us also choose $m_\rho = \frac{7}{3}m_\phi$. This choice is by no means unique, any value satisfying the limits $2m_\phi < m_\rho < 4m_\phi$ would do as well.

**(a)** $\vec{P}=0$ sector: the $\phi$-particles have opposite momenta, the smallest value is $p_\phi = (2\pi/L)$, and $p_\rho = 0$:

$$W_\rho = m_\rho\,; \qquad W_{2\phi} = 2\sqrt{m_\phi^2 + p_\phi^2}$$

Using the "would-be resonance" condition $W_\rho = W_{2\phi}$, we see that the required lattice size is $L\approx 10.5/m_\phi$.

**(b)** $\vec{P}\neq 0$: now we can choose $p_{\phi,1} = (2\pi/L)$ and $p_{\phi,2}=0$, so that $p_\rho = (2\pi/L)$ and

$$W_\rho = \sqrt{m_\rho^2 + p_\rho^2}\,; \qquad W_{2\phi} = m_\phi + \sqrt{m_\phi^2 + p_{\phi,1}^2}\,.$$

This gives $L \approx 4.5/m_\phi$ at the resonance. The ratio of volumes in cases (a) and (b) is $\frac{V_a}{V_b} = \left(\frac{10.5}{4.5}\right)^3 \approx 12.7$! The non-zero momentum sector gains more than an order of magnitude over the zero momentum case.

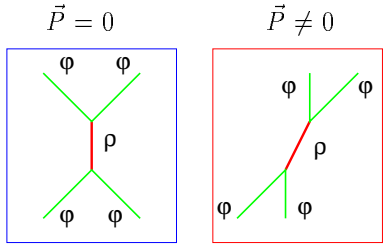

Even in $\vec{P} \neq 0$ case it is convenient to work in the the center of mass frame of the scattering particles. In this frame the 2-particle energy is, in terms of the lattice (laboratory) frame energy $W_L$,

$$W_{\mathrm{cm},2\phi} = \sqrt{W_{L,2\phi}^2 - \vec{P}^2} \qquad (1)$$

and the momenta of the particles

$$p_{\mathrm{cm},1} = p_{\mathrm{cm},2} = \sqrt{\tfrac{1}{2}W_{\mathrm{cm},2\phi} - m_\phi}. \qquad (2)$$

In fig. 1, we show the 2-particle center of mass frame energy spectrum $W_{2\phi}$ both for $\vec{P} = 0$ and $\vec{P} = 2\pi/L$ sectors. The lattice sizes calculated in the previous example are indicated by arrows. When we turn the (small) $\rho\phi\phi$ interaction on, the points where the $2\phi$ levels cross the $\rho$ levels turn into *avoided level crossings*, and $\rho$ becomes unstable.

## 2. PHASE SHIFT

The phase shift $\delta_l$ parametrizes the partial wave decomposition of the scattering amplitude:

$$T = \frac{16\pi W_{2\phi}}{2ip} \sum_{l=0}^{\infty} (2l+1) P_l(\cos\theta)(e^{2i\delta_l} - 1). \qquad (3)$$

In 3+1 dimensions the connection between the 2-particle energy levels and $\delta$ is quite involved, and we present only the final formula here [6]. For concreteness, let $\vec{P} = 2\pi\hat{x}/L$. The energy $W_{\mathrm{cm},2\phi}$ is measured, and the CM-frame momentum $p_{\mathrm{cm}}$ is calculated from

$$p_{\mathrm{cm}} = \sqrt{W_{\mathrm{cm},2\phi}^2 - m_\phi^2}. \qquad (4)$$

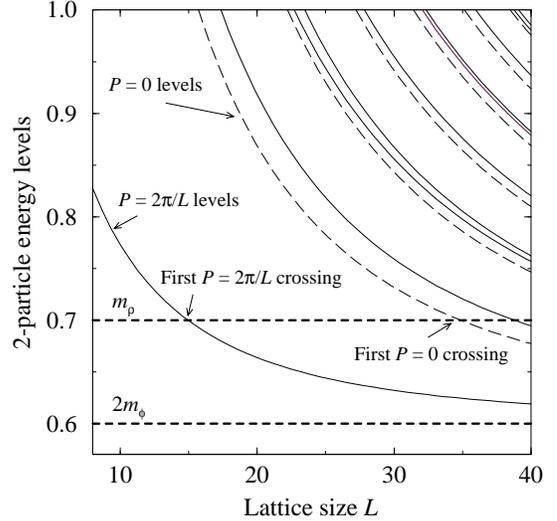

Figure 1. The free particle center of mass energy levels for the $\vec{P} = 0$ (dashed lines) and $\vec{P} = 2\pi/L$ (solid lines) sectors.

In the s-wave ($l = 0$) sector, the relation between $p_{\mathrm{cm}}$ and the phase shift $\delta_0$ is given by

$$\delta_0(p_{\mathrm{cm}}) = -\Phi(q) \bmod \pi, \qquad q = \frac{p_{\mathrm{cm}} L}{2\pi}, \qquad (5)$$

where $\Phi$ is a continuous function defined by

$$\tan(-\Phi(q)) = \frac{\gamma q \pi^{3/2}}{Z_{00}(1; q^2)} \qquad \Phi(0) = 0, \qquad (6)$$

with $\gamma = (1-v^2)^{-1/2} = (1 + \vec{P}^2/W_{\mathrm{cm}}^2)^{1/2}$. Function $Z_{00}$ is a generalized zeta function, given by

$$Z_{00}(s; q^2) = \frac{1}{\sqrt{4\pi}} \sum_{\vec{r} \in A} (\vec{r}^2 - q^2)^{-s}, \qquad (7)$$

where the set $A = \{\vec{r} \in \mathbf{R}^3 | r_1 = \gamma^{-1}(n_1 + 1/2), r_2 = n_2, r_3 = n_3; \vec{n} \in \mathbf{Z}^3\}$. The expansion of $Z_{00}$ is convergent when $\mathrm{Re}\, s > 3/2$, but it can be analytically continued to $s = 1$ [6]. When $\vec{P} \to 0$, $\gamma \to 1$ and $A \to \mathbf{Z}^3$, and Lüscher's original formula [1] is recovered.

## 3. SIMULATIONS AND RESULTS

We test the method with a simple 3+1 dimensional spin model containing two coupled Ising spins

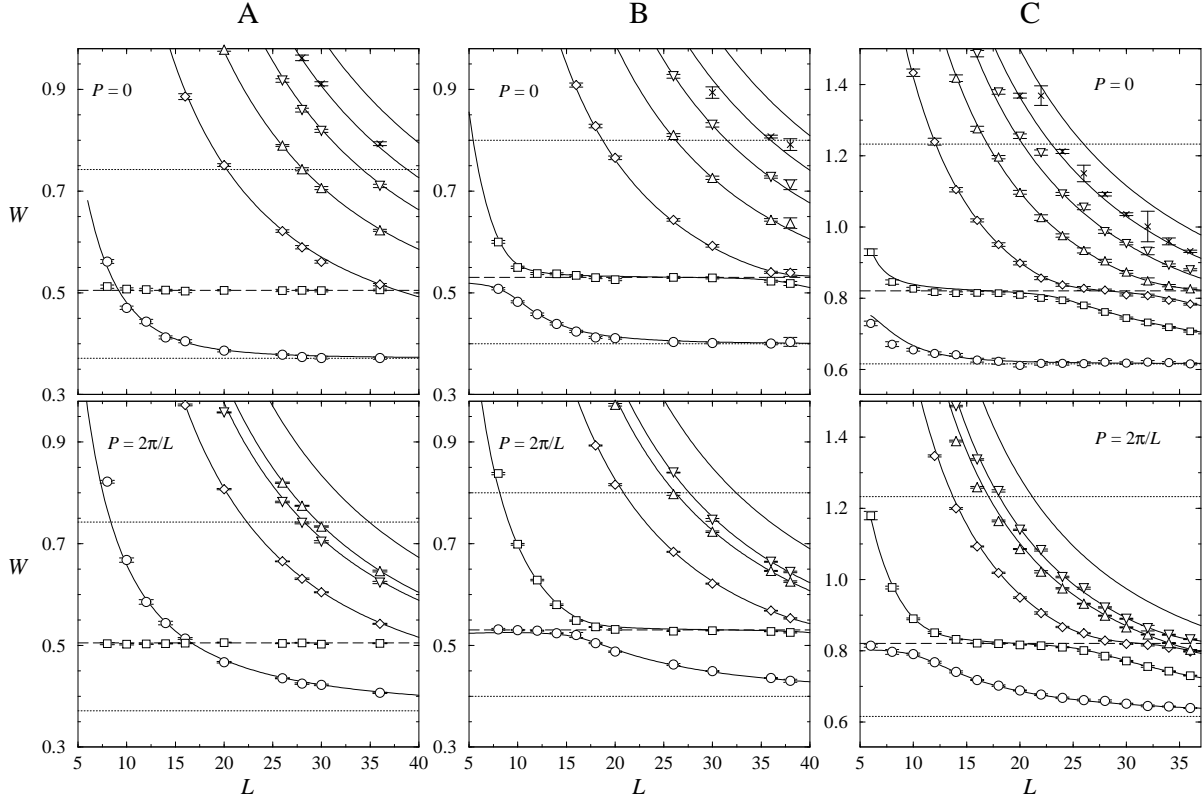

Figure 2. The center of mass energy levels for sectors $\vec{P} = 0$ (top row) and $\vec{P} = 2\pi/L$ (bottom) for cases A, B and C (see table 1).

$$S = \sum_{\langle x,y \rangle}[-\kappa_\phi \phi_x \phi_y - \kappa_\rho \rho_x \rho_y + g\rho_x \phi_x \phi_y], \quad (8)$$

where $\phi, \rho = \pm 1$. Choosing the coupling constants suitably $\rho$ appears as a resonance in s-wave $\phi\phi$ scattering. Let us now define operators

$$\mathcal{O}(\vec{P}, \vec{p}; t) = \tilde{\phi}(\vec{P} + \vec{p}; t)\tilde{\phi}(-\vec{p}; t), \quad (9)$$

where $\tilde{\phi}$ is the spatial Fourier transform of $\phi$. We calculate the phase shift as follows: (i) we measure the cross-correlation matrices of operators $\mathcal{O}$ for both $\vec{P} = 0$ and $\vec{P} = 2\pi/L$ separately; (ii) the exponential falloff of the eigenvalues of the matrices yields the energy levels; (iii) using eqs. (4–7) we obtain $\delta_0(p_{\rm cm})$ [6].

We use 3 sets of couplings (A, B, C), given in table 1. In fig. 2 we show the center of mass energy levels for both of the $\vec{P}$ sectors. In case A the coupling $g = 0$ and $\rho$ is stable; in cases B and C $g \neq 0$ and the crossings of the $\rho$ and $2\phi$ levels turn into avoided level crossings. In the $\vec{P} = 0$ sector even the lowest $2\phi$ energy level increases when $L$ decreases. This is due to the repulsive $\lambda\phi^4$ interaction in the scaling limit of the Ising model, and in cases A and B it causes the level crossing at $L \approx 8$. However, these points are plaqued by severe finite volume effects, and the first useful crossing occurs only around $L = 37$. This shifts down to $L \approx 16$ when $\vec{P} \neq 0$.

Table 1
The couplings of runs A, B and C, and the measured scattering parameters.

|  | A | B | C |
|---|---|---|---|
| $\kappa_\phi$ | 0.0742 | 0.07325 | 0.07075 |
| $\kappa_\rho$ | 0.0708 | 0.0718 | 0.0665 |
| $g$ | 0 | 0.008 | 0.021 |
| $m_\phi a$ | 0.1856(4) | 0.1996(5) | 0.3081(4) |
| $m_\rho a$ | 0.5049(5) | 0.5306(13) | 0.8206(11) |
| $\Gamma a$ | 0 | 0.0044(2) | 0.0178(7) |
| $g_R a$ | 0 | 0.598(14) | 1.49(3) |
| $\lambda_R$ | 28.1(1.1) | 36.8(1.3) | 48.3(2.0) |

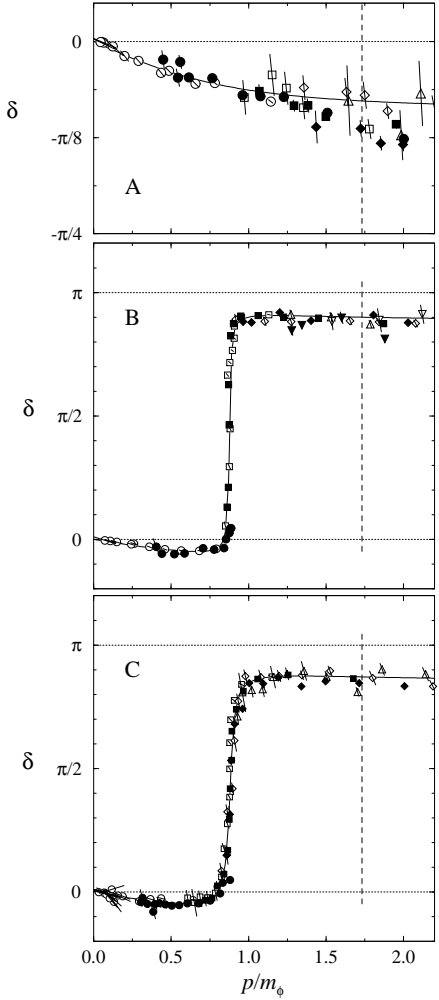

Figure 3. The phase shifts for cases A ($g = 0$), B (0.008) and C (0.021). Filled and open symbols correspond to $\vec{P} = 0$ and $2\pi/L$, respectively.

In fig.3 we show the phase shifts for $\vec{P} = 0$ and $2\pi/L$ for A, B and C. The repulsive $\lambda\phi^4$ interaction is evidenced by the negative slope ($\propto$ scattering length) of $\delta$ near $p = 0$. In B and C the resonance appears as a rapid increase of $\delta$ near $W_{\text{cm}} = m_\rho$. The width of the resonance is proportional to the slope of $\delta$ at $\pi/2$. Note the excellent consistency of $\vec{P} = 0$ and $2\pi/L$ sectors (filled and open symbols).

The continuous lines in fig.3 are perturbative fits to the $\delta_0$ data:

$$\delta_0^{\text{pert}} = -\lambda_R \frac{p}{16\pi W} + \frac{g_R^2}{32\pi} \frac{1}{Wp} \log \frac{4p^2 + m_\rho^2}{m_\rho^2}$$
$$- \tan^{-1}\left[\frac{g_R^2}{16\pi} \frac{p}{W} \frac{1}{W^2 - m_\rho^2}\right]. \qquad (10)$$

The fitted parameters $\lambda_R$, $g_R$ and $m_\rho$ are listed in table 1. Inverting eqs. (4–7) we obtain the energy levels corresponding to $\delta^{\text{pert}}$, as shown in fig.2 with continuous lines. Note that *all* the levels are fitted simultaneously. Finally, the resonance width is calculated from

$$\Gamma_\rho = \frac{g_R^2}{32\pi m_\rho^2} \sqrt{m_\rho^2 - 4m_\phi^2}. \qquad (11)$$

## Conclusions

We have measured resonance scattering parameters to a high accuracy using Monte Carlo methods in 3+1 dimensions. The sectors $\vec{P} = 0$ and $2\pi/L$ are completely compatible within statistical errors; however, level crossings occur in the $\vec{P} = 2\pi/L$ sector in much smaller volumes. The sectors complement each other: for a fixed volume, they give $\delta(p)$ with different argument $p$; and both sectors can be calculated simultaneously, often at little extra cost.